\documentclass[aps,prl,twocolumn,superscriptaddress,groupedaddress,nofootinbib]{revtex4}  
\usepackage{graphicx}  
\usepackage{dcolumn}   
\usepackage{bm}        
\usepackage{amssymb}   
\newcommand{\st}{\scriptsize}
\usepackage{bm}
\usepackage[usenames ,dvipsnames]{xcolor}
\usepackage{slashed}
\usepackage{lipsum}
\usepackage{graphicx,color}
\usepackage{url}
\usepackage{stfloats}
\usepackage{dsfont}        
\usepackage{slashed}
\usepackage{epsfig}
\usepackage{multirow,array}
\usepackage{float}
\usepackage{fancyhdr}
\usepackage{indentfirst}
\usepackage{cancel}
\usepackage{tabularx}
\usepackage{simplewick}
\usepackage{amsmath}
\RequirePackage[colorlinks=true
  ,urlcolor=Green
  ,anchorcolor=blue
  ,citecolor=orange
  ,filecolor=blue
  ,linkcolor=blue
  ,menucolor=blue
  ,pagecolor=blue
  ,linktocpage=true
  ,pdfproducer=medialab
  ,pdfa=true
]{hyperref}

\begin{document}


\title{Stochastic Gravitational Wave Background from Global Cosmic Strings}
     
\author{Chia-Feng Chang}
\email{chiafeng.chang@email.ucr.edu}
\author{Yanou Cui} 
\email{yanou.cui@ucr.edu}                                                      
\affiliation{Department of Physics and Astronomy, University of California, Riverside, CA 92521, USA}

\date{\today}

\begin{abstract}

Global cosmic strings are generically predicted in particle physics beyond the Standard Model, e.g., a post-inflationary global $U(1)$ symmetry breaking which may associate with axion-like dark matter. We demonstrate that although subdominant to Goldstone emission, gravitational waves (GWs) radiated from global strings can be observable with current/future GW detectors. The frequency spectrum of such GWs is also shown to be a powerful tool to probe the Hubble expansion rate of the Universe at times prior to the Big Bang nucleosynthesis where the standard cosmology has yet to be tested. 
\end{abstract}
\maketitle

\section{Introduction}
The recent discovery of gravitational waves (GW) by the LIGO/Virgo collaboration \cite{Abbott:2016,Aasi:2014,TheLIGOScientific:2017,TheVirgo:2014} ushered in a new era of observational astronomy, and offered opportunities to probe new fundamental physics. Among potential cosmological sources for a stochastic GW background (SGWB) \cite{Caprini:2018mtu}, a cosmic string network is one that can generate strong signals over a wide frequency range \cite{Hindmarsh:1994re, Vilenkin:2000jqa}, and is among the primary targets for SGWB searches at GW experiments such as LIGO and LISA \cite{Abbott:2017mem, Auclair:2019wcv}. Cosmic strings are stable one-dimensional objects characterized by a tension $\mu$. They can arise from superstring theory \cite{Copeland:2003bj, Dvali:2003zj} or from a vortex-like solution of field theory \cite{Nielsen:1973cs} which typically descends from a spontaneously broken gauge or global $U(1)$ symmetry. Field theory string tension relates to the symmetry breaking scale $\eta$ by $\mu\propto\eta^2$ while for global strings there is an additional divergence factor $\sim\ln(\eta/H)$ ($H^{-1}$: horizon size) \cite{Copeland:1990qu, Dabholkar:1989ju}. Once formed, the network consists of horizon-size long strings along with a collection of sub-horizon sized string loops due to long string intersections. The loops subsequently oscillate and radiate energy until they decay away. 

For many types of cosmic strings such as Nambu-Goto (NG) strings, GW is the dominant radiation mode. In contrast, GWs from global strings have been largely neglected as it is in general subdominant to Goldstone boson or axion emission (only a few attempts exist \cite{Battye:1993jv, Battye:1996pr, Battye:1997ji, Figueroa:2012kw, Ramberg:2019dgi}). However, since the detection prospect of the Goldstones are highly model-dependent, the GW signal, albeit rare, could be complementary or even a smoking-gun for discovery in the case when the Goldstones/axions have no non-gravitational interaction with the Standard Model (SM) (e.g. \cite{Arvanitaki:2009fg}). It is particularly timely to re-examine this overlooked signal channel, in light of the growing interest in axion-like dark matter models where axion strings are inevitably present for post-inflationary $U(1)_{PQ}$ breaking.

In this \textit{Letter}, we perform a state-of-the-art calculation for GW frequency spectrum originated from global strings and demonstrate that such GW signals can be observable with current/future GW detectors, while consistent with existing constraints. A comparison with NG string generated GWs is also made. Furthermore, we investigate the effects of non-standard background cosmology on this spectrum. We thereby demonstrate that, analogous to (yet distinct from) the case with NG or gauge strings \cite{Cui:2017, Cui:2018, Cui:2019kkd, Dror:2019syi, Gouttenoire:2019rtn, Gouttenoire:2019kij}, global string induced GW spectrum can provide a new window to probe beyond the Standard Model particle physics as well as be a powerful tool to discern the energy composition of the Universe during the pre-Big Bang nucleosynthesis (BBN) \textit{primordial dark age} \cite{Boyle:2007zx, Boyle:2005se}. 
\section{Gravitational Wave Spectrum from Global Cosmic Strings}
Results from the simulations for Abelian-Higgs string or NG string network have demonstrated that after formation the network quickly reaches a scaling regime, and GW radiation is the leading energy loss mechanism \cite{Vachaspati:1984yi, Hindmarsh:1994re, Olum:1999sg, Vilenkin:2000jqa, Hindmarsh:2017qff, Matsunami:2019fss}\footnote{Many studies including the very recent \cite{Matsunami:2019fss} have reached consensus of GW domination for Abelian-Higgs strings, while a few studies differ, e.g. \cite{Hindmarsh:2017qff}. As suggested in \cite{Blanco-Pillado:2017oxo, Matsunami:2019fss} the discrepancy in \cite{Hindmarsh:2017qff} could be due to insufficient resolution used in simulation on cosmological scales, and that Abelian-Higgs strings are expected to approach NG limit on such large scales.}. In contrast, simulation for global strings is much more challenging due to the need to cover a large hierarchy in the relevant physical scales: the string core size $\sim \eta^{-1}$ and the inter-string separation scale $H^{-1}$. Recent years have seen rapid development in global string simulations \cite{Klaer:2017qhr,Gorghetto:2018myk,Kawasaki:2018bzv,Hindmarsh:2019csc,Buschmann:2019icd} partly driven by its connection to axion physics, while uncertainties remain to be resolved with future higher resolution simulations, in particular whether there is a logarithmic deviation from scaling \cite{Gorghetto:2018myk, Buschmann:2019icd, Hindmarsh:2019csc} \footnote{The discrepancy among different simulation results could be due to different numerical algorithms and diagnostics for counting strings \cite{Martins:2018dqg, Buschmann:2019icd}, in addition to the different ranges of excess tension $N$ they explored \cite{Buschmann:2019icd}.}.

\subsection{Scaling Solution and the VOS Model}
As in \cite{Martins:2018dqg}, we take the available simulation results at face value for our studies. To integrate the string network evolution into our studies on GW signal, we adopt the approach in \cite{Martins:1996jp,Martins:2000cs, Martins:2018dqg} which is based on the analytical velocity-dependent one-scale (VOS) model. We focus on oscillating string loops, expected to be the leading sources of both GW and Goldstone emissions \cite{Vilenkin:2000jqa,Battye:1997jk}. Given the present uncertainties on global string simulation, we will first consider a simple monochromatic loop size distribution at formation time $\ell_i=\alpha t_i$. Inspired by the recent NG string simulations \cite{Blanco-Pillado:2013qja,Blanco-Pillado:2017oxo}, we consider a benchmark scenario that $10\%$ of the network energy releases to $\alpha\simeq0.1$ large loops while the remaining goes to kinetic energy of smaller loops that dissipates by redshifts. Later we will also present results for a log uniform distribution up to $\alpha\sim1$ as suggested in \cite{Gorghetto:2018myk} based on simulating the first few e-folds of Hubble expansion after the formation of a global string network. 

Global strings are characterized by a time-dependent string tension \cite{Copeland:1990qu, Dabholkar:1989ju, Vilenkin:2000jqa}
\begin{equation}
\mu(t) = 2 \pi \eta^2 \hbox{ln}\left(L/\delta\right) \equiv 2 \pi \eta^2 N, \label{eq: tension}
\end{equation}
where $L\simeq H^{-1}\xi^{-1/2}$ is the string correlation length, $\xi$ is the number of long strings per horizon volume, $\delta\simeq1/\eta$ is the string thickness, and $N\equiv\hbox{ln}\left(L/\delta\right) \simeq \hbox{ln}(\eta\xi^{-1}t)$ is time-dependent. The evolution equations for the correlation length $L$ in a global cosmic string network are \cite{Vilenkin:2000jqa,Martins:2018dqg}:
\begin{eqnarray}
\label{Eq:stringEoM}
\left( 2 - \frac{1}{N} \right) \frac{dL}{dt} &=& 2 H L \left( 1 + \bar{v}^2 \right) + \bar{c} \bar{v} + s \frac{\bar{v}^6}{N},\\
\frac{d\bar{v}}{dt} &=& \left( 1 - \bar{v}^2 \right) \left[ \frac{k}{L} - 2 H \bar{v}\right],
\end{eqnarray}
where $k$ is a momentum parameter.  The terms on the RHS of Eq.~\ref{Eq:stringEoM} represent free expansion, loop chopping rate and Goldstone radiation back-reaction, in order. The average long string velocity $\bar{v}$, the loop chopping parameter $\bar{c}$ and Goldstone radiation parameter $s$ ($\sigma\equiv s\bar{v}_0^5$ will be used later to quantify such radiation loss) will be determined by calibrating with recent simulation results. By treating the Goldstone radiation as a perturbative process as in \cite{Martins:2018dqg} we can solve the above equations starting with the standard scaling solution without the radiative corrections (denoted with subscript $0$), and obtain solutions of the following form:
\begin{eqnarray}
\left(\frac{L}{t}\right)^2&=&\left(\frac{L}{t} \right)^2_0(1+\Delta)\\
\bar{v}^2&=&\bar{v}_0^2(1-\Delta),
\end{eqnarray}
where $\Delta$ represents the radiative correction. We extend the analysis in \cite{Martins:2018dqg} to generic cosmology background parametrized by $n$ (background energy density $\rho\propto a^{-n}$ where $a$ is the scale factor), and derive the number of strings per Hubble volume $\xi$ and velocity in the scaling regime:
\begin{align}
\label{Eq:VOS}
\xi & = \frac{8 \left(1 - \frac{2}{n} - \frac{1}{2N} \right)}{n k (k + \bar{c})(1+\Delta)}, \\ \bar{v}^2 & = \frac{n - 2 - \frac{n}{2N}}{2} \frac{k}{k + \bar{c}} (1 - \Delta),
\end{align}
with
\begin{align}
\Delta \equiv \frac{s}{N(k+\bar{c})} \left[ \frac{n - 2 - \frac{n}{2N}}{2} \frac{k}{k + \bar{c}}\right]^{5/2},
\end{align} 
The results from field theory simulations for global topological defects (summarized in Table.~\ref{Table1}) provide calibration points of $\xi, \bar{v}$ for certain choices of $N$. Our $\chi^2$ fitting obtain the best fits with $\sim 3.3\sigma$ significance, leading to the calibrated model parameters $\{ \bar{c}, k, \sigma \} \simeq \{ 0.50, 0.28, 5.83 \}$. The error bars in Table.~\ref{Table1} are extracted from \cite{Klaer:2017ond,Gorghetto:2018myk,Hindmarsh:2019csc,Kawasaki:2018bzv}. We only include the data with $N > 5$ in our $\chi^2$ fitting, since the VOS model may not be accurate in the very low $N$ regime. It is worth mentioning that a persisting linear growth of $\xi$ in $N$, as suggested by extrapolating simulation results based on very early evolution \cite{Gorghetto:2018myk, Buschmann:2019icd}, cannot be reproduced in the VOS model, in agreement with \cite{Martins:2018dqg}.

\begin{table}[H]
\centering
\begin{tabular}{l| ccc}
  \hline
    \hline
  \hbox{Reference} & $N$ & $\xi$ & $\bar{v}$ \\
  \hline
\hbox{Klaer et al.} \cite{Klaer:2017ond} &$ 55$ &$ 4.4 \pm 0.4$ & $0.50 \pm 0.04$ \\
 &$ 31$ &$ 4.0 \pm 0.4$ & $0.50 \pm 0.04$ \\
  &$ 15$ &$ 2.9 \pm 0.3$ & $0.51 \pm 0.04$ \\
 \hbox{Gorghetto et al.} \cite{Gorghetto:2018myk} &$ 6-7$ &$ 1.0 \pm 0.30$ &  \\
 \hbox{Hindmarsh et al.} \cite{Hindmarsh:2019csc} &$ 6$ &$ 1.2 \pm 0.20$ &  \\ 
  \hbox{Kawasaki et al.} \cite{Kawasaki:2018bzv} &$ 2-4$ &$ 1.1 \pm 0.30$ &$ 0.52 \pm 0.05$  \\ 
    \hline
      \hline
\end{tabular}
\caption{Results of recent global string network simulations for the number of strings per Hubble volume, $\xi$, and the average velocity of long strings $\bar{v}$. This table contains data points used in \cite{Martins:2018dqg} and an additional one from the more recent \cite{Hindmarsh:2019csc}. \label{Table1}}
\end{table}%
Once reaching the scaling regime, the long string energy density evolves as
\begin{equation}
\rho_\infty = \xi(t) \frac{\mu(t)}{t^2},
\end{equation}
where $\xi(t)$ quickly approaches a constant for NG strings, yet needs to be determined for global strings. We find that the analytical VOS model predicts the evolution of physical parameters $\xi\sim 4.0$, $\bar{v}\sim 0.57$ at times deep into the radiation domination (RD) era $(N \gtrsim 20)$.

\begin{figure}[H]
\centering
\includegraphics[width=0.44\textwidth]{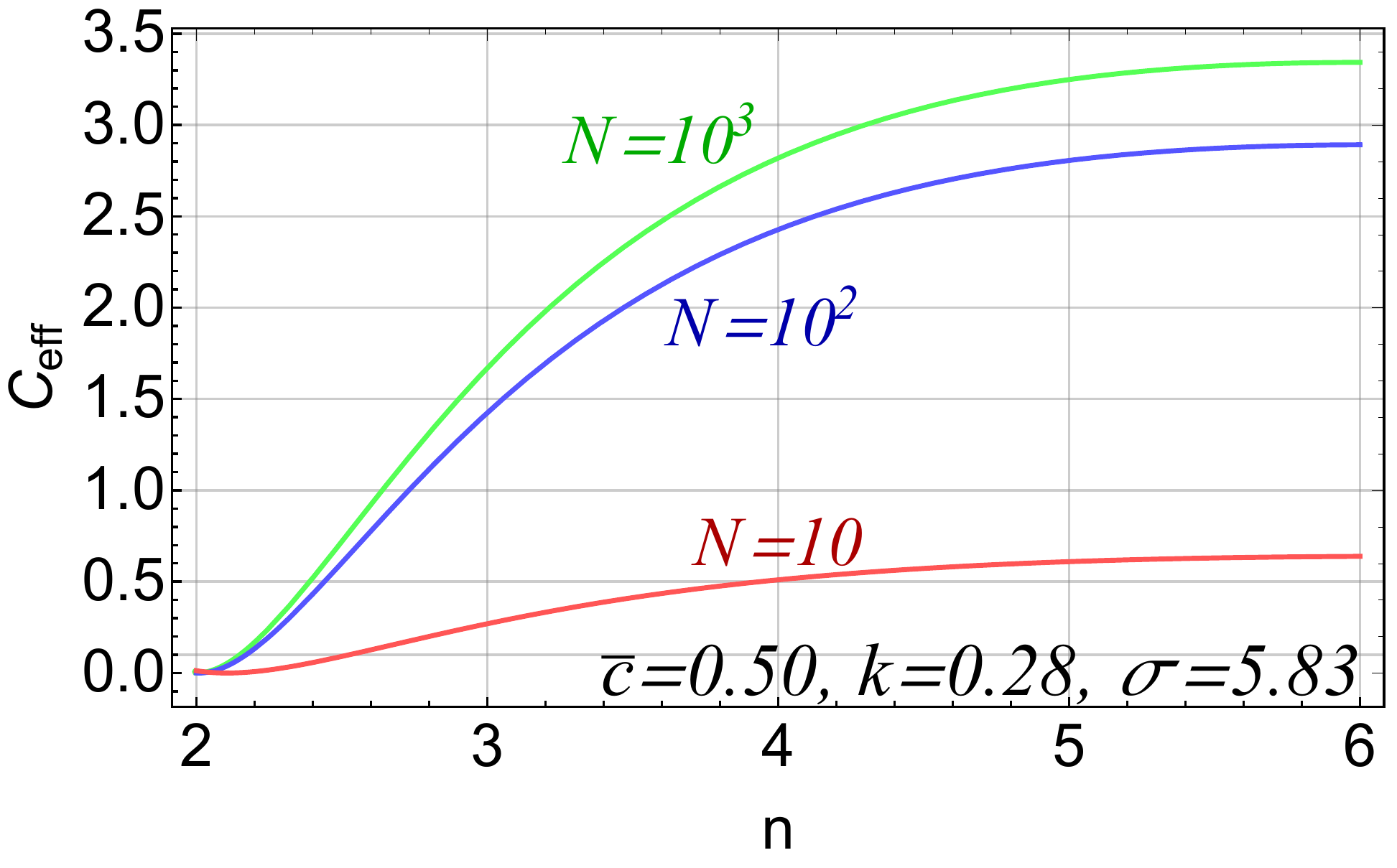} 
\caption{\label{FS1} Dependence of the loop formation rate parameter $C_{\hbox{\st{eff}}}$ on $N$ (time parameter) and background cosmology model (parametrized by $n$), derived from the VOS model.}
\end{figure}

Applying local energy conservation, the loop formation rate per unit volume at the formation time $t_i$ is given by
\begin{equation}
\frac{d n _{\hbox{\st{loop}}}}{dt_i} \simeq F_{\alpha} \frac{C_{\hbox{\st{eff}}}(t_i)}{\alpha} t_i^{-4}. \label{eq: loop_formrate}
\end{equation}
As said, we will first consider the NG string motivated simple case of $F_\alpha\sim10\%,~\alpha\sim 0.1$. The relation of $C_{\hbox{\st{eff}}}(t_i) \simeq \xi^{3/2}\bar{c}\bar{v}$ can be predicted based on the VOS model solutions and depends on the redshift scaling of the background energy density $\rho$. As shown in Fig.~\ref{FS1}, for $n=3$ (matter domination), 4 (radiation domination), and 6 (kination), we find $C_{\hbox{\st{eff}}} \simeq 1.32, 2.26, 2.70 $ at large $N$, respectively. 

\subsection{Radiations from a Global String Network}
Once formed, a loop oscillates and loses energy by the rate \cite{Vilenkin:1986ku, Battye:1995hw, Battye:1993jv, Vilenkin:2000jqa}
\begin{equation}
dE/dt=-\Gamma G\mu^2-\Gamma_a\eta^2, \label{eq: powers}
\end{equation} where the right hand side represents GW and Goldstone radiation in order. Studies show that the dimensionless constants $\Gamma\simeq 50$ \cite{Vilenkin:1981bx,BlancoPillado:2011dq,Blanco-Pillado:2013qja,Blanco-Pillado:2017oxo}, $\Gamma_a\simeq 65$ (by employing the antisymmetric tensor formalism for radiation from explicit loop solutions \cite{Vilenkin:2000jqa, Vilenkin:1986ku} or applying energy conservation in VOS model \cite{futurework}). Note that considering the $2\pi\hbox{ln}\left(L/\delta\right)$ factor in $\mu$ (Eq.~\ref{eq: tension}), GW and Goldstone radiation rates can be comparable at late times for sufficiently large $\eta$ ($\gtrsim10^{15}$ GeV). For simplicity we ignore the radiation of heavy radial modes which is suppressed relative to Goldstone emission \cite{Gorghetto:2018myk}, and would not noticeably affect the GW signal.

Consequently the length of a loop after its formation time $t_i$ would evolve as
\begin{equation}
\ell(t) \simeq \alpha t_i - \Gamma G \mu (t-t_i) - \kappa (t-t_i), \label{eq: ell_evol}
\end{equation}
where $\kappa \equiv \Gamma_a/(2\pi N)$.

\begin{figure}[H]
\centering
\includegraphics[width=0.48\textwidth]{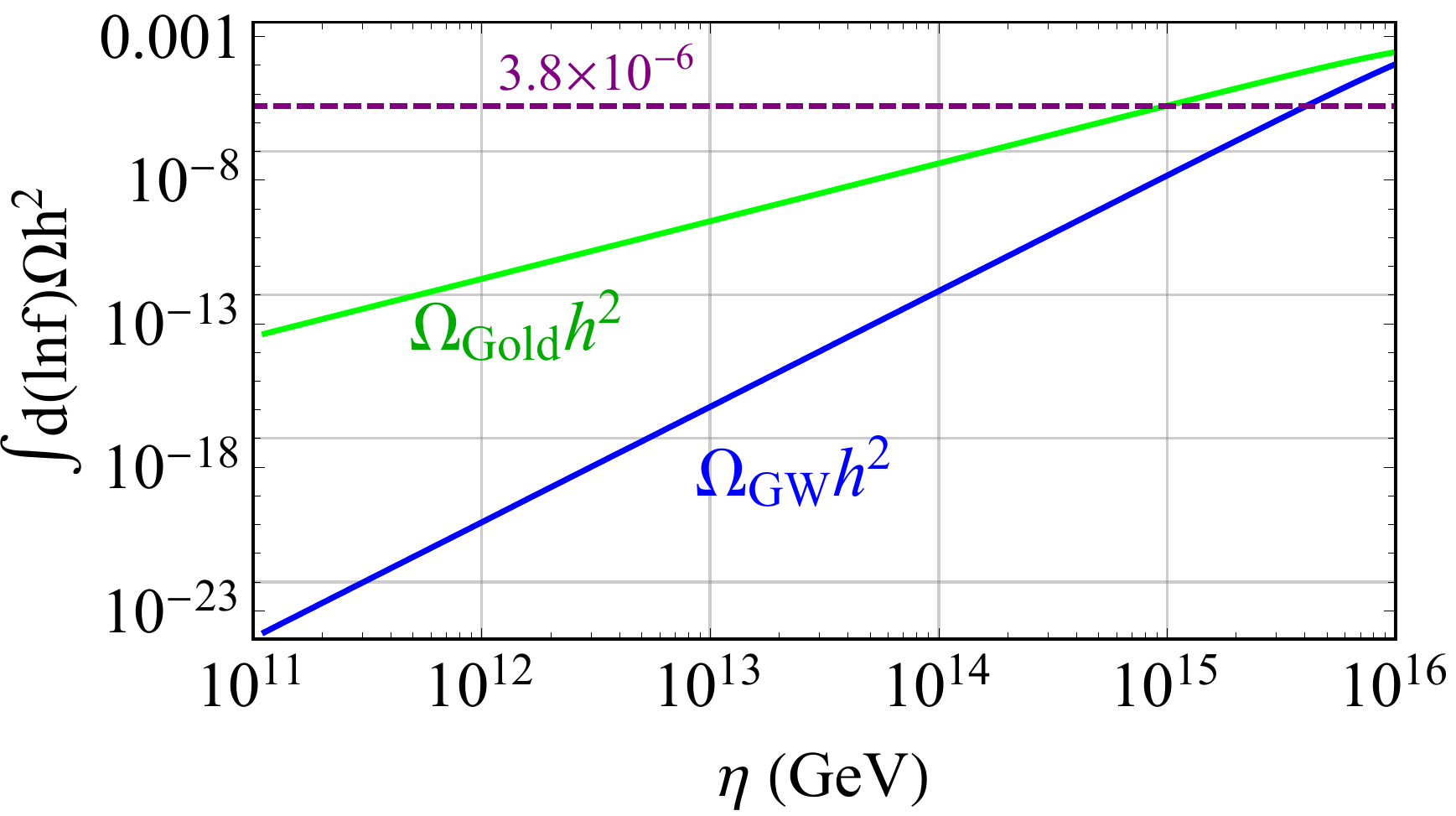} 
\caption{\label{FS2}  The relic densities of massless Goldstone boson (green) and gravitational waves (blue) from global cosmic strings with varying symmetry breaking scale $\eta$. The purple dashed line indicates the constraint on extra radiation from the CMB data.}
\end{figure}

The string loops emit GWs from normal mode oscillations at frequencies $f_{\hbox{\st{emit}}} \simeq 2k/\ell$ with $k \in \mathds{Z}^+$. The emitted GW frequencies then redshift as 
\begin{equation}
\label{eq: f_redshift}
f \simeq \frac{a(\tilde{t})}{a(t_0)} \frac{2k}{\ell(\tilde{t})},
\end{equation}
where $\tilde{t}$ is GW emission time, $t_0$ is the current time. Summing up contributions from all harmonic modes and using 
Eqs.~\ref{eq: loop_formrate}, \ref{eq: ell_evol}, the GWs relic density spectrum as observed today is
\begin{equation}
\label{eq: GWspect_f}
\Omega_{\hbox{\st{GW}}} (f) = \frac{f}{\rho_c} \frac{d \rho_{\hbox{\st{GW}}}}{df} = \sum_k \Omega_{\hbox{\st{GW}}}^{(k)} (f).
\end{equation}
with 
\begin{align}
\label{eq: GWspect1}
\notag \Omega_{\hbox{\st{GW}}}^{(k)} (f) = & \frac{1}{\rho_c} \frac{2k}{ f} \frac{\mathcal{F_\alpha}}{\alpha} \int^{t_0}_{t_F} d\tilde{t} ~\Theta(t_i ,\tilde{t})\frac{\Gamma^{(k)} G \mu^2}{\alpha + \Gamma G \mu + \kappa} \\& \times  \frac{C_{\hbox{\st{eff}}}\left( t_i^{(k)} \right)}{t_i^{(k)4}} \left( \frac{a(\tilde{t})}{a(t_0)} \right)^{5} \left( \frac{a(t_i^{(k)})}{a(\tilde{t})} \right)^{3},
\end{align} 
where $t_F$ is the formation time of the string network, $\rho_c = 3 H_0^2 /8\pi G$ is the critical density, and the decomposed radiation constant as $\Gamma^{(k)} \equiv \Gamma k^{-\frac{4}{3}}/3.6$, with the causality and energy conversation conditions $\Theta(t_i,\tilde{t}) \equiv \theta(\tilde{\ell}) \theta(\tilde{t}-t_i)$. To consider radiation of Goldstones, we can define $\Omega_a(f)$ in analogy to Eq.~\ref{eq: GWspect1}, and simply replace $\Gamma\rightarrow\Gamma_a$, $\Gamma G\mu^2\rightarrow\Gamma_a\eta^2$. 

\begin{figure}[H]
\includegraphics[width=0.48\textwidth]{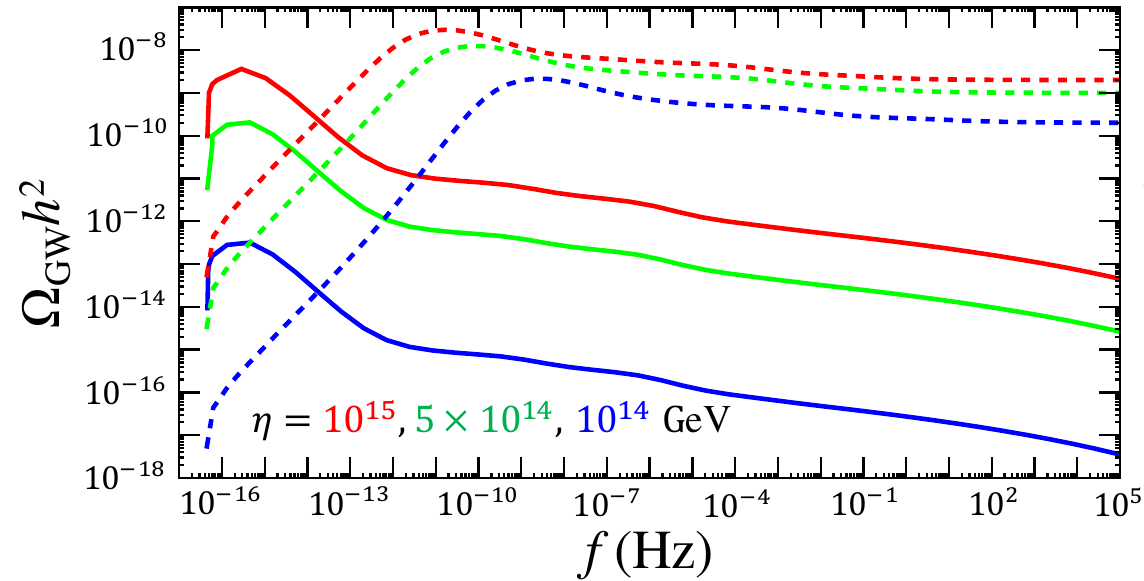} 
\caption{\label{F1} Gravitational wave frequency spectrum (standard cosmology) from a global (solid) vs. NG (dashed) string network with $\alpha = 0.1$ and $\eta = 10^{15}$ (red),\,$5\times 10^{14}$ (green),\, $10^{14}$ (blue) GeV.}
\end{figure}

In order to understand the relative importance of GW vs. Goldstone emission, it is illuminating to estimate and compare the relic densities of the two modes integrated over frequencies. We define the integrated energy densities as
\begin{align}
\Omega_\beta = \int d \left( \hbox{ln}f \right) \Omega_\beta (f), \;\;\;\;\;\; \hbox{with} \;\; \beta = \{ \hbox{GW}, \; \hbox{Gold} \}, 
\end{align} 
where $\Omega_\beta (f)$ is estimated based on Eq.(\ref{eq: GWspect_f}). Fig.\ref{FS2} illustrates the relative importance of $\Omega_{\rm GW}$ and $\Omega_{\rm Goldstone}$ with dependence on $\eta$. We can see that the relative suppression of GW vs. Goldstone radiation is less severe at larger $\eta$, which contributes to rendering detectable GW signals from global strings with $\eta\gtrsim 10^{14}$ GeV.

\subsection{Results}

We are now ready to demonstrate the GW spectra in benchmark scenarios. To determine the viable range of $\eta$, we first take into account the self-consistency condition that the $U(1)$ symmetry breaking occurs or is restored after inflation \cite{Hertzberg:2008,Gibbons:1977,Bunch:1978, Bassett:2005} which requires $\eta \lesssim 1.2\times 10^{16}$GeV based on recent CMB data \cite{Aghanim:2018,Akrami:2018,Array:2015xqh}. Therefore we will consider $\eta\lesssim O(10^{15})\,$GeV in our benchmark examples.

In Fig.~\ref{F1}, assuming standard cosmic history, we show the GW frequency spectrum originated from global cosmic strings with varying $\eta$. The NG string GW spectra \cite{Cui:2017,Cui:2018} are also shown for comparison. One can see that the global string GW amplitudes are more sensitive to $\eta$: $\Omega^{\rm global}_{\rm GW}\propto \eta^4$ vs. $\Omega^{\rm NG}_{\rm GW}\propto \eta$ \cite{futurework}. We also observe that, relative to NG string spectrum, the global string GW spectrum overall shifts to lower frequency, and the magnitude of the shift depends on $\eta$. This can be explained by the key relationship between the loop lifetime $\tau$ and $\eta$ by solving Eq.(\ref{eq: ell_evol}):
\begin{equation}
\tau \simeq \frac{\alpha + \Gamma G \mu + \kappa}{\Gamma G \mu +\kappa} t_i,
\end{equation}
where the NG string scenario is restored with $\kappa = 0$. We can see that due to the strong Goldstone emission rate, global string loops typically decay within one Hubble time, i.e., have a much shorter lifetime than their NG string counterpart. Because of this, radiation from global strings on average experiences a longer period of redshift, rendering a spectrum shifted towards lower frequency. Meanwhile, the logarithmic time dependence of $\mu$ made a gradually, logarithmically declining plateau towards high $f$ on the GW spectrum, instead of a nearly flat plateau in NG strings. Such a logarithmic tail towards high $f$ is a unique feature for the GW spectrum from global strings that can distinguish it from other sources of stochastic GW background. For instance, an astrophysical foreground can arise from a collection of weaker, unresolved binary mergers, but the spectrum rises by $f^{2/3}$ power-law towards high $f$ and peaks around $f\sim10^3$ Hz \cite{TheLIGOScientific:2016dpb}. The GW spectrum from NG strings are nearly flat towards high $f$ range, and a summary about the spectral shapes of other cosmological sources of GWs can be found in e.g. \cite{Cui:2018}, which all differ from the case with global strings.

\begin{figure}[h]
\centering
\includegraphics[width=0.43\textwidth]{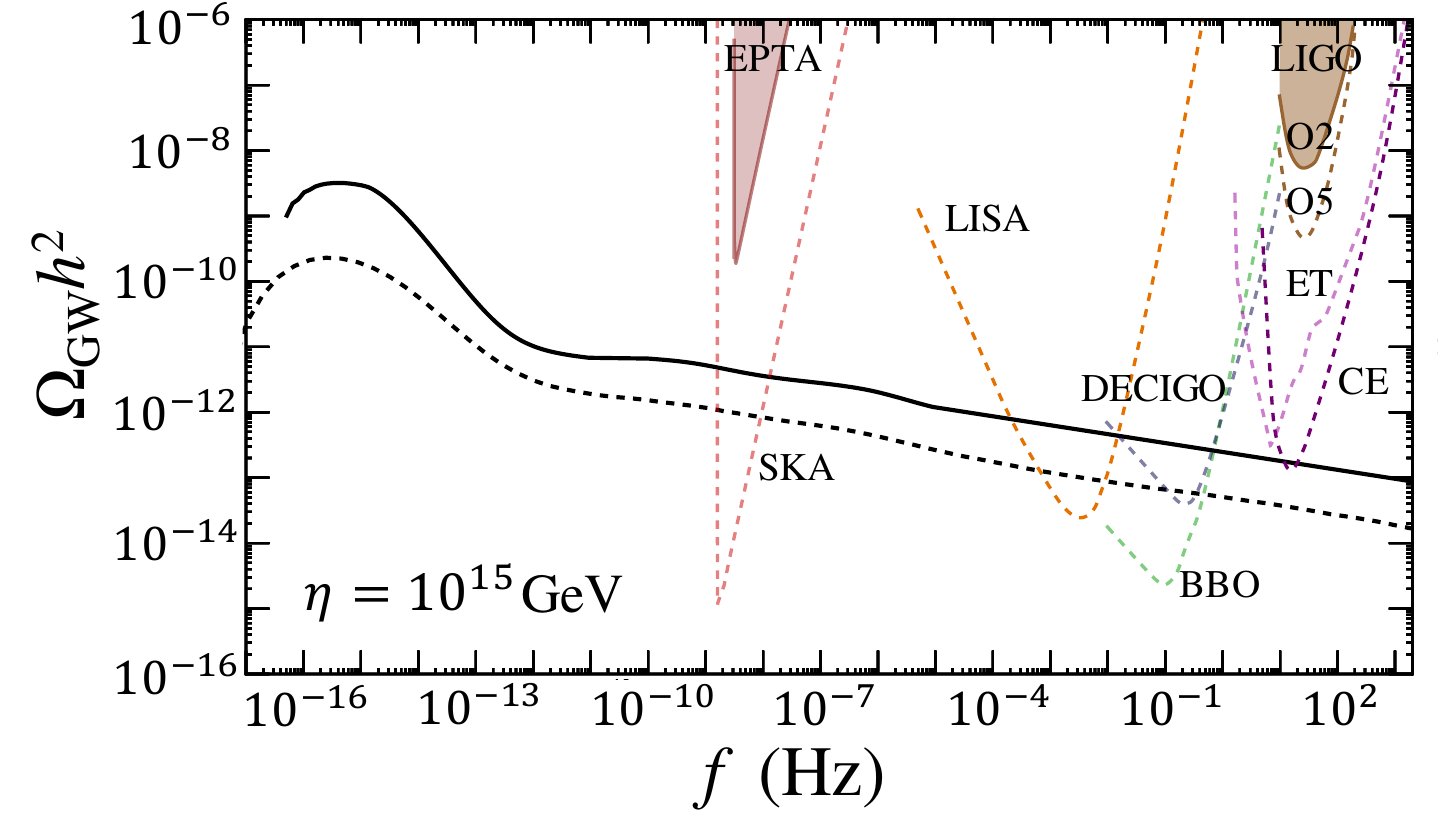} 
\caption{\label{F3} Gravitational wave frequency spectrum from a global string network with $\eta = 10^{15}\,$GeV and $\alpha = 0.1$ assuming a standard cosmological history (solid black). The spectrum at frequencies lower (higher) than $10^{-13}\,$Hz corresponds to matter (radiation) dominated era. The dashed black spectrum line is based on a log uniform loop size distribution ($\alpha \leq 1$) at formation as suggested by a recent simulation \cite{Gorghetto:2018myk}. Upper shaded regions/dashed lines show current constraints/future sensitivities with various GW experiments.}
\end{figure}

In Fig.~\ref{F3}, a benchmark example of a GW frequency spectrum is given with standard cosmological history. For comparison, we also show the results based on an alternative scenario where sub-horizon loops are formed with a logarithmic uniform distribution \cite{Gorghetto:2018myk}. We can see that albeit a slight difference in amplitude, the GW spectrum is mostly robust against such an uncertainty in loop distribution. The spectrum based on log uniform distribution extends to slightly lower $f$ due to the formation of larger loops closer to horizon size. We also checked the effect on $\Omega_{\hbox{\st{GW}}}(f)$ by varying the loop emission parameter $\Gamma_a$, and found that approximately $\Omega_{\hbox{\st{GW}}}(f) \propto \Gamma/\Gamma_a$, which can be estimated by energy conservation law based on Eq.(\ref{eq: powers}). In the limit of $\Gamma_a\ll \Gamma$, $\Omega_{\hbox{\st{GW}}}(f)$ is no longer sensitive to $\Gamma_a$ as GW radiation dominates.

 We also show the current sensitivity bands of LIGO \cite{Abbott:2016, Aasi:2014, Thrane:2013oya,LIGOScientific:2019vic} and the projected sensitivities for LISA \cite{Bartolo:2016ami}, DECIGO/BBO \cite{Yagi:2011}, Einstein Telescope (ET) \cite{Punturo:2010zz,Hild:2010id} and Cosmic Explorer (CE) \cite{Evans:2016mbw}. In the lower $f$ region, the European Pulsar Timing Array (EPTA) \cite{vanHaasteren:2011ni} imposes a strong constraint of $\eta \lesssim 3.2 \times 10^{15}\,$GeV, with the expected sensitivity of SKA shown below \cite{Janssen:2014dka}.

 Additional constraints come from the potential distortion of the CMB power spectrum by global strings \cite{Ade:2015xua, Charnock:2016, Lopez-Eiguren:2017dmc} and the CMB/BBN bound on the total energy densities of GWs and radiation-like Goldstones \cite{Smith:2006nka,Henrot-Versille:2014jua,Tanabashi:2018}, which lead to $\eta \lesssim 10^{15}\,$GeV. In general CMB polarization data potentially yields a strong bound of $\Omega_{\rm GW}h^2 \lesssim 10^{-11}$ in the range $f\sim 10^{-17}-10^{-15}$ Hz \cite{Lasky:2015lej,Smith:2005,Namikawa:2019tax}. However, GWs from global strings safely evade this bound since the GW signal in this very low frequency range is not populated until after the photon decoupling, thus is not present at the CMB epoch. To be specific, numerically we find that the GWs from global strings in the $f\sim 10^{-17}-10^{-15}$ Hz band are emitted when $T\sim10^{-4}-10^{-3}$ eV (can also be estimated based on Eq.(\ref{eq: f-T})), which is well after the photon decoupling temperature $T_{\gamma} \sim 0.3\,$eV.
 
\indent In summary, Fig.~\ref{F3} demonstrates that SGWB from global strings can be observable with foreseeable GW detectors while satisfying existing constraints. This is in contrast to some earlier simulation-based studies \cite{Lopez-Eiguren:2017dmc, Fenu:2009qf, Figueroa:2012kw} which drew a pessimistic conclusion. Although it takes time to fully resolve the issue, a reasonable understanding is that the discrepancy originates from the fact that simulations cannot capture dynamics at larger string separation scales (or later times in the string network evolution) beyond their current limitation (up to $\sim1000$ times of the string core size, or $N\sim10$), while our semi-analytical approach covers the full time range of interest \footnote{A simple illustration of this explanation can be made by examining the lifetime of a loop of size $L$ based on Eqs.~\ref{eq: tension}, \ref{eq: powers}. With $N\sim10$ we find $\tau\sim L$, agreeing with the recent simulation result \cite{Saurabh:2020pqe} and the general observation from simulations that the loop decays before it has time to effectively oscillate and produce GWs. However, on much larger scales of cosmological relevance, $\tau\propto N$, e.g., with $N\sim100$, $\tau\sim10L$, allowing sufficient oscillations to produce observable GWs. Our studies show that the loop dynamics on such large scales provides the leading contribution to GWs which is beyond the regime that the current simulation can cover.}. 
\section{The Effects of non-standard cosmologies}
Relic GW spectrum from cosmic strings is influenced by an extended period of cosmic history (Eq.~\ref{eq: GWspect1}) and consequently populates signals over a wide range of $f$. As a result, such a spectrum can be used to test standard cosmology and probe potential deviations prior to the BBN epoch. This idea has been investigated in the context of NG strings \cite{Cui:2017, Cui:2018, Caldwell:2018giq}. Here we demonstrate the results for global strings, which are analogous to, yet distinct from, the NG string case. \\
\begin{figure}[h]
\centering
\includegraphics[width=0.45\textwidth]{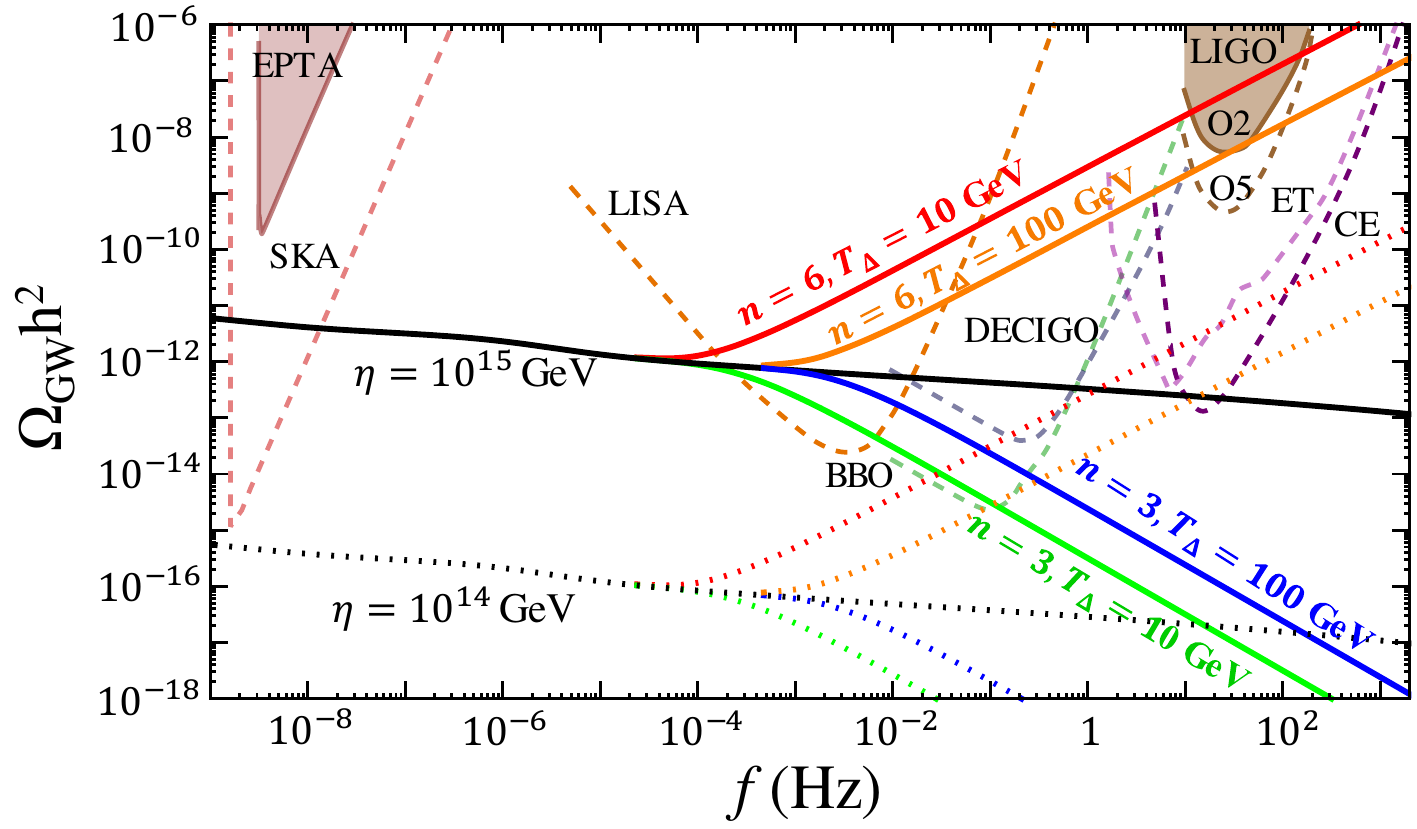} 
\caption{\label{F5} Gravitational wave frequency spectrum from a global cosmic string network with $\alpha = 0.1$ with $\eta = 10^{15}\,$GeV (solid lines) and $\eta=10^{14}\,$GeV (dotted lines). The colored (black) lines show the GW spectrum with the non-standard (standard) cosmological evolution. The related experimental sensitivities are also shown.}
\end{figure}\\
\indent We consider two types of well-motivated non-standard histories. The first is a period of early matter domination ($n=3$) before the onset of the standard RD era. This can be due to a temporary energy dominance by a long-lived massive particle or an oscillating scalar moduli field in a quadratic potential \cite{Moroi:1999zb}. The second modification to standard cosmology is a period of kination, with $n>4$, which can arise from the oscillation of scalar field in non-renormalizable potential in quintessence models for dark energy or inflation \cite{Salati:2002md, Chung:2007vz}. In particular, these scenarios have been considered in recent work on axion cosmology \cite{Poulin:2018dzj, Nelson:2018via, Ramberg:2019dgi}. \\
\indent We define the $t_\Delta(T_{\Delta})$ as the time (radiation temperature) when the  Universe transits to standard RD. The evolution of the energy density of the Universe before and after such a transition can then be parametrized as
\begin{align}
\rho(t) = \left\{            
\begin{aligned}
&\rho_{\hbox{\st{st}}}(t_\Delta)\left[ \frac{a(t_\Delta)}{a(t)} \right]^n \;\;\; ; t< t_\Delta \\
&\rho_{\hbox{\st{st}}}(t) \;\;\;\;\;\;\;\;\;\;\;\;\;\;\;\;\;\;\;\;\;\; ; t\geq t_\Delta
\end{aligned}
\right.
\end{align}
where $\rho_{\hbox{\st{st}}}$ is the energy density assuming standard cosmology, and $n=3$ $(n=6)$ for early matter (kination) domination. In addition, to be consistent with BBN \cite{Hannestad:2004px}, in both scenarios the transition temperature should satisfy $T_{\Delta} \gtrsim 5\,$MeV.
In Fig.~\ref{F5} we demonstrate the effects of non-standard cosmologies on the GW spectrum from global strings, along with experimental sensitivities. Benchmark parameters are: $\eta=10^{15}, 10^{14}$ GeV, and $T_{\Delta}=10,100$ GeV. One can see dramatic deviations from the standard prediction: a distinct falling (rising) at high $f$ spectrum due to the transition to an early matter-dominated (kination) phase. LIGO data already excluded the kination case with $T_{\Delta} \lesssim 20\,$GeV, $\eta \gtrsim 10^{15}\,$GeV. Potential overproduction of GWs and radiation-like Goldstones imposes additional constraint on the kination case. We have checked that other cases shown in Fig.~\ref{F5} are consistent with these limits. Furthermore, these bounds can be alleviated depending on the onset/duration of kination domination \cite{futurework}.
\begin{figure}[h]
\centering
\includegraphics[width=0.44\textwidth]{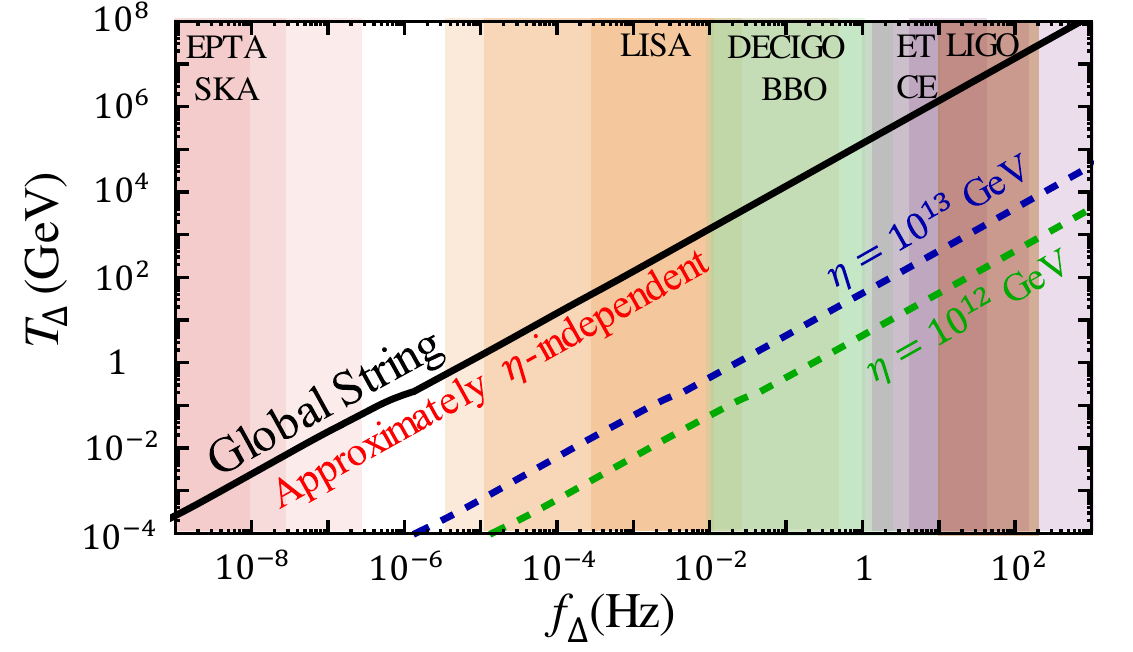} 
\caption{\label{F4} Frequency $f_\Delta$ where the GW spectrum from cosmic strings would be altered due to the transition to a non-standard cosmology at $T_\Delta$ (Eq.~\ref{eq: f-T}): the comparison of the results for global strings (the upper-left black line) vs. NG strings \cite{Cui:2017,Cui:2018} (the lower-right dashed lines) ($\alpha=0.1$). The relevant experimental sensitivities are also shown where darker bands correspond to peak sensitivities.}
\end{figure}

As shown in Fig.~\ref{F5}, the transition to a non-standard cosmology at $T_\Delta$ results in a deviation in the GW spectrum at certain frequency $f_\Delta$. Similar to the case with NG strings \cite{Cui:2017,Cui:2018}, at frequencies above $f_\Delta$ the spectrum rises following a power-law of $f^{+1}$ for kination or falls as $f^{-1}$ for early MD. The correspondence between characteristic GW frequency and radiation temperature in the RD era can be derived rigorously, analogous to the method for the NG string case \cite{futurework}, or approximately with Eq.~\ref{eq: f_redshift} (considering $k\simeq1$ mode dominance \cite{Hagmann:1990mj, Vilenkin:2000jqa}) based on the observation that the global string loops are short-lived. The analytical estimate agrees with the numerical result:
\begin{align}
\label{eq: f-T}
f_\Delta \simeq 3.02 \times 10^{-6}\, \hbox{Hz} \, \left( \frac{{T_\Delta}}{1\, \hbox{GeV}} \right) \left( \frac{\alpha}{0.1} \right)^{-1} \left[ \frac{g_*(T_\Delta)}{g_*(T_{\hbox{\st{eq}}})} \right]^{1/4}.
\end{align}
In Fig.~\ref{F4} we illustrate the above relation and the comparison with its counterpart for NG strings \cite{Cui:2017,Cui:2018}. We can see that, unlike in the NG string case, $\eta$-dependence is nearly absent in the $f_\Delta-T_\Delta$ relation (also as shown in Eq.~\ref{eq: f-T}), and with the same $f_\Delta$ band, GWs from global strings generally probe higher $T_\Delta$ or earlier times. These differences originate from the shorter lifetime of global string loops. As a result, provided that the signal amplitude is sufficient, with detectors such as ET/CE GWs from global strings can probe cosmology up to a very early epoch of $T_\Delta \sim 10^{8}\,$GeV, even well before the reach of NG strings of $T_\Delta\sim 10^{4}\,$GeV \cite{Cui:2017} (for NG strings, $\eta\gtrsim10^{13}$ GeV is excluded by EPTA assuming standard cosmology). Meanwhile, as shown in Fig.~\ref{F4} for global strings the latest possible transition at $T_\Delta\sim5$ MeV is within the coverage of SKA/EPTA.

\section{Conclusion and Discussion}
Stochastic GW background (SGWB) from a global cosmic string network is a universal signature and potential discovery mode for the underlying theory typically involving a global $U(1)$ symmetry breaking which may associate with axion-like dark matter physics. In this work we compute the frequency spectrum originated from such a GW source based on recent analytical modeling and numerical simulation for global/axion strings. We demonstrate that, despite the apparent dominance of Goldstone radiation over GWs, global string generated SGWB can be within reach of an array of current/planned GW detectors, while compatible with existing constraints. We have also shown how such a GW spectrum depends on the equation of state of the background cosmology, and thereby can be utilized to probe the early Universe far before the BBN epoch which would be unaccessible with other observational means.

Although this work focuses on global strings associated with massless, radiation-like Goldstones, it may also lead to a new avenue for probing massive axion-like particle (ALP) dark matter models. For axion models with post-inflationary $U(1)_{\rm PQ}$ breaking, axion topological defects including strings and domain walls are indispensable companions to axion dark matter particles. Recently there have been a substantially increased interest in studying axion topological defects \cite{Klaer:2017qhr, Ferrer:2018uiu, Gorghetto:2018myk, Saikawa:2017hiv, Long:2019lwl, Buschmann:2019icd, Hindmarsh:2019csc}. Given that the detection strategy for axion particles is highly model-dependent, the universal GW signals from axion topological defects can be highly complementary or even the smoking-gun, in particular for hidden axion models \footnote{Other recent examples of probing hidden axions with stochastic GW background include e.g. \cite{Machado:2019xuc, Machado:2018nqk} (these are not sourced by cosmic strings). }. Due to the non-zero mass of axions and the related late-time formation of domain walls, the calculation of the GW spectrum in ALP models is more complex. We will explore this direction in future work \cite{futurework}, which can be an important complement to the existing literature on ALP models.

\begin{acknowledgments}
\noindent \textit{{\textbf{Acknowledgments.}}}
We thank Edward Hardy, Hitoshi Murayama, Thomas Schwetz-Mangold and Wei Xue for helpful discussions. We also thank Daniel Figueroa, Mark Hindmarsh, Marek Lewicki, Joanes Lizarraga, Asier Lopez Eiguren, David Morrissey and Jon Urrestilla for commenting on the manuscript. The authors are supported in part by the US Department of Energy grant DE-SC0008541. CC thanks Yin Chin Foundation of U.S.A. for its support. YC thanks the Kavli Institute for Theoretical Physics (supported by the National Science Foundation under Grant No. NSF PHY-1748958), Erwin Schr\"odinger International Institute, and Galileo Galilei Institute for hospitality while the work was being completed.
\end{acknowledgments}

\bibliographystyle{apsrev4-1}
\bibliography{References}

\end{document}